\documentclass[letterpaper, 10 pt, conference]{ieeeconf}  

\IEEEoverridecommandlockouts                              

\usepackage{amsmath} 
\usepackage{amssymb}  
\usepackage{graphicx}
\usepackage{booktabs}
\usepackage{algorithmicx}
\usepackage{algorithm}
\usepackage{algpseudocode}

\usepackage{todonotes} 

\algdef{SE}[SUBALG]{Indent}{EndIndent}{}{\algorithmicend\ }%
\algtext*{Indent}
\algtext*{EndIndent}

\title{\LARGE \bf
Koopman-based Differentiable Predictive Control for the Dynamics-Aware Economic Dispatch Problem
}

\author{Ethan King, J\'an Drgo\v na, Aaron Tuor, 
Shrirang Abhyankar, Craig Bakker, Arnab Bhattacharya, Draguna Vrabie

\thanks{This research is part of the Data Model Convergence Initiative at Pacific Northwest National Laboratory. It was conducted under the Laboratory Directed Research and Development Program at PNNL, a multiprogram national laboratory operated by Battelle for the U.S. Department of Energy.}
\thanks{The authors are with the Pacific Northwest National Laboratory, Richland, Washington, USA
         \{ethan.king, jan.drgona, aaron.tuor, shrirang.abhyankar, craig.bakker, arnab.bhattacharya,  draguna.vrabie\}@pnnl.gov }%
}

\begin{document}

\maketitle
\thispagestyle{empty}
\pagestyle{empty}

\begin{abstract}
The dynamics-aware economic dispatch (DED) problem embeds low-level generator dynamics and operational constraints to enable near real-time scheduling of generation units in a power network. DED produces a more dynamic supervisory control policy than traditional economic dispatch (T-ED) that leads to reduced overall generation costs. However, the incorporation of differential equations that govern the system dynamics makes DED an optimization problem that is computationally prohibitive to solve. In this work, we present a new data-driven approach based on  differentiable programming to efficiently obtain parametric solutions to the underlying DED problem.
In particular, we employ the recently proposed differentiable predictive control (DPC) for offline learning of explicit neural control policies using an identified Koopman operator (KO) model of the power system dynamics. We demonstrate the high solution quality and five orders of magnitude computational-time savings of the DPC method over the original online optimization-based DED approach on a 9-bus test power grid network.
\end{abstract}

\section{INTRODUCTION}

Power grid operators solve the economic dispatch (ED) problem to determine the power output at generation units that will meet system loads at minimal cost subject to operational constraints. Traditionally ED solves for supervisory set-points under steady state assumptions every 5 minutes given look-ahead forecasts of loads~\cite{wood2013}. Varying real-time demand is then met by lower-level Automatic Generation Control (AGC) that dynamically adjusts generator output around the ED set-points to account for changing loads~\cite{thatte2011}. The ED optimization is solved under assumptions that the system frequency is at a stable nominal value (60 Hz in the US), valid for high inertia synchronous generators and slow-varying loads, while the real-time AGC actions ensure that system frequency is maintained within safe limits~\cite{thatte2011,chakraborty2020}. 

An increasing number of renewable generation sources in the grid and changing demand-response mechanisms can result in large frequency fluctuations and can destabilize conventional synchronous operations~\cite{milano2018}. Current ED approaches may become less valid, placing a greater burden on AGC to maintain system frequency within safe limits while requiring larger and more frequent deviations from ED set-points, which can lead to sub-optimal power generation costs~\cite{khatami20}. There is a need for ED approaches that account for the generator and frequency dynamics in dispatch policies to alleviate the reliance on AGC corrections. 

Methods to integrate ED and AGC have been explored; including frequency aware economic dispatch~\cite{thatte2011,lee2013}, model predictive control approaches~\cite{Khler2017RealTE}, and multi-objective optimization methods~\cite{li2016, trip2016}. In an approach developed in ~\cite{khatami20} and ~\cite{chakraborty2020} the separation of ED and AGC is maintained, but the ED set-points are improved by the incorporation of generator dynamics into the ED optimization. The method provides continuous dynamic ED set-points that require less AGC corrections. The authors in~\cite{khatami20} show that this dynamics-aware ED (DED) approach can produce more economical operation and reduce variance in system dynamics in comparison to traditional ED. The approach also has the additional benefit of being easily implementable within existing system architectures.

In general, generator dynamics are nonlinear and their introduction into ED results in a nonlinear differential-algebraic optimization problem that is computationally challenging to solve. The authors in~\cite{khatami20} reduce DED to a linear programming problem using a function space-based method with an aggregate model of the system dynamics. An alternative approach to simplifying DED is taken in~\cite{king2021} using a Koopman operator (KO) framework to construct a linear representation of the generator dynamics. For a full review of state-of-the-art methods for incorporating these dynamics see~\cite{Shri2017}. While the methods introduced in ~\cite{khatami20} and~\cite{king2021} reduce the computational complexity of DED, the optimization can still be prohibitively expensive to solve online. For the method in~\cite{king2021}, on a small 9-bus example, the solution time was over a quarter of the forecast horizon.

Fast solution at scale is required to implement DED effectively, and high computational overhead may limit its effectiveness, particularly during periods with large fluctuations in frequency where repeated solutions over a forecast horizon could be of significant benefit. Here we explore the use of an offline optimization approach based on differential predictive control (DPC)~\cite{drgona2020differentiable} to offer substantial speedups at solution time. 
DPC is a newly developed method for offline learning of constrained neural control policies by sampling the state space and computing policy gradients based on sensitivities of the computational graph representing constraints of the underlying model predictive control (MPC) problem~\cite{drgona2022Gurantees}.
In particular, the DPC approach leverages differentiable programming~\cite{DiffProg2019} to obtain the gradients of the MPC objectives with respect to the control policy parameters subject to differentiable system dynamics as well as state and input constraints.
Other works explore the potential of differentiating through the implicit MPC policy to optimize some of the MPC parameters (e.g., system dynamics models, constraints bounds, or weight matrices) to improve the performance or robustness in online optimization settings~\cite{diffMPC2018,east2020infinitehorizon,safe_RL_MPC_2019}.
For practical applications, it has been shown that learning control policies by differentiating the MPC problem is a  computationally efficient offline solution suitable for building control problems~\cite{DRGONA202114,GNURL2019}.
Others have addressed the limitations of computing explicit MPC policies offline, including deep learning to approximate the control policies given training data computed by an online MPC~\cite{maddalena2019neural,Hertneck8371312,Chen2018,KARG2021107266,DRGONA2018}. However, the disadvantage of approximate MPC approaches is the reliance on the online solutions of the original MPC problem. In contrast, DPC represents an unsupervised approach for approximating explicit MPC policies without the need for online solvers.

In this work, we demonstrate the applicability of the DPC approach for obtaining explicit solutions of the DED problem with significant computational speedups at inference time. A major computational challenge to using a DPC approach for DED is the repeated simulations of the nonlinear swing dynamics in a differentiable architecture that is necessary for training a neural control policy. We reduce the computational burden of learning the control by utilizing a linear KO model of the generator dynamics from~\cite{king2021} in the DPC computational graph. The learned DPC control policy for DED is an open loop control parameterized by the load-forecasts and generation costs over a long time horizon, which constitutes learning a mapping between high dimensional inputs and outputs. We achieve this mapping by downsampling the input load-forecasts and using convolutional neural network layers to construct the control policy. DPC involves higher offline costs to learn the parametric map but we show can provide several orders of magnitude speedups at execution.

\section{Dynamics-Aware Economic Dispatch}
To compute generator policies for a power grid the DED problem relies on a model of the grid and generator dynamics. In this work we use a simplified model given by the swing equations as presented in~\cite{Arapostathis82}.

\subsection{Swing Dynamics} 

The swing equations are a nonlinear system of differential equations which capture the generator mechanical dynamics. In this paper, we also assume the presence of a swing bus operating at a fixed synchronous frequency with voltage angle $\theta_{s}$ and a fixed angular frequency $\omega_{s}$. Within a grid network of generator nodes $\mathcal{G}$ and load nodes $\mathcal{L}$ we denote the angle of the voltage at each node $i$ at time $t$ relative to the swing bus by $\theta_{i}(t)$. Each generator $i\in \mathcal{G}$ is assumed to have dynamics governed by the swing equation  
\begin{equation}
M_{i}\ddot{\theta}_{i}(t)+ D_{i}\dot{\theta}_{i}(t) = P_{i}(t) - f_{i}(\theta(t)) \, ,
\label{swng_gen}
\end{equation}
where $M_{i}$ is the moment of inertia, $D_{i}$ is the damping constant, and $P_{i}(t)$ is the mechanical input torque to the generator at time $t$. The load nodes $i\in \mathcal{L}$ are assumed to satisfy the power flow equation
\begin{equation}
L_{i}(t) - f_{i}(\theta(t)) = 0 \, .
\end{equation}
For both the generator and load nodes, the amount of power injected into the network at load $i$ is given by
\begin{equation}
f_{i}(\theta(t)) = \sum_{j} |V_{i}||V_{j}| |Y_{i,j}| \sin(\theta_{i}(t) - \theta_{j}(t)) \, ,
\label{swng_f}
\end{equation}
where with $|V_{i}|$ the magnitude of the voltage at each node and $Y_{i,j}$ the admittance of the line between $i$ and $j$. The voltages are assumed to be given and fixed at all nodes.

\subsection{Dynamics-Aware Economic Dispatch Formulation}

Given a forecast of loads on the grid, to solve the DED problem the operator must provide generator inputs $P_{i}(t)$  to satisfy those loads at minimal cost over the forecast horizon while maintaining the system within operational constraints. This is a nonlinear non-convex differential-algebraic optimization problem. Discretizing the differential equations with a given numerical integration scheme (forward Euler, for example) for a given time horizon and time-discretization $\mathcal{T}$, the resultant formulation for DED problem is

{\small
\begin{subequations}
\begin{align}
&\min_{P_{i}(t_{k}) \,  i\in \mathcal{G}, k\in \mathcal{T} } \sum_{k \in \mathcal{T} } \left( \sum_{i \in \mathcal{G}} c_{i} P_{i}(t_{k})  + c_{s} |P_{s}(t_{k}) | \right) \, ,  \\
&\mbox{subject to $~ \forall ~ k \in \mathcal{T}$ :} \nonumber\\
&M_{i}\dot{\omega}_{i}(t_{k}) + D_{i}\omega_{i}(t_{k}) = P_{i}(t_{k}) - f_{i}(\theta(t_{k}))  , ~ \forall ~  i\in \mathcal{G} \cup \{s\} \label{swng1} \\
&\omega_{i}(t_{k}) = \dot{ \theta}_{i}(t_{k})\, , ~ \forall ~   i\in \mathcal{G} \cup \{s\} \, , \\
&\omega_{s}(t_{k}) = 0 \, , \\
&P_{i}(t_{k}) - f_{i}(\theta(t_{k})) = 0  \, , ~ \forall ~   i\in \mathcal{L} \, , \label{swng4} \\
&P_{i}^{\min} \leq P_{i}(t_{k})  \leq P_{i}^{\max}\, , ~ \forall ~  i\in \mathcal{G} \, , \label{gbnd}\\
&\omega_{i}^{\min} \leq \omega_i(t_{k})  \leq \omega_{i}^{\max} \, , ~ \forall ~  i \in \mathcal{G} \, , \label{wbnd} \\
&  | P_{i}(t_{k+1}) - P_{i}(t_{k}) | \leq \epsilon_{rmp}\, , ~ \forall ~  i \in \mathcal{G} \,  \label{genrmp}  \\
& P_{i}(t_{0} ) = P_{0_{i}} ~  \forall ~ \, i \in \mathcal{G} \label{g_init}
\end{align}
\label{DED_OG}
\end{subequations}
}
The objective captures the total cost of generation at cost $c_{i}$ for generator $i$ and $c_{s}$ for the slack bus. Note that if total generation exceeds the total load on the system the value of the slack bus will become negative. Negative slack bus values are equally penalized by the absolute value term in the objective. The constraints \eqref{swng1}--\eqref{swng4} capture the swing dynamics. Operational constraints are captured by \eqref{gbnd} - \eqref{g_init}, where \eqref{gbnd} bounds generator inputs, \eqref{wbnd} bounds generator frequencies, \eqref{genrmp} bounds the change in the generator input allowed over each time step by $\epsilon_{rmp}$ and \eqref{g_init} enforces the initial condition of the generators.

The swing dynamics constraints \eqref{swng1}--\eqref{swng4} make this a computationally challenging optimization problem, however they can be simplified greatly by using a Koopman Operator (KO) approach as done in~\cite{king2021}.

\subsection{Model Reduction with the Koopman Operator}

The KO captures a systems dynamics with respect to functions of its state called observables. Critically, observables can be constructed such that their evolution in time, given by the KO, is linear. The approach can be thought of as lifting nonlinear dynamics to linear dynamics on a space of observables. In general, the KO is infinite dimensional, but in practice system dynamics can be estimated with a finite dimensional KO approximation. An overview of the KO and additional technical details can be found in~\cite{bakker20cp,budisic12jsr}.   

We use the KO model developed in~\cite{king2021} to capture the swing dynamics. Let $x_{k}$ the system state at time $k$, consisting of the phase angles, slack bus value, and generator frequencies. The observables are taken to be the vector of generator frequencies $\boldsymbol \omega_{k} $, the slack bus value $P_{s_{k}}$, and the output of a multilayer perceptron (MLP) $N(x_{k};w)$ with weights $w$, giving the vector of observables
\begin{equation}
\psi(x_{k}) =  \begin{pmatrix}  \boldsymbol \omega_{k} \\ P_{s_{k}} \\ N(x_{k};w) \end{pmatrix} \, .
\end{equation}
The generator frequencies and slack bus value are included in the observables such that they can be easily recovered to compute the objective and evaluate the DED state constraints, while the outputs of $N(x_{k};w)$ are trained such that the swing dynamics can be closely approximated. 

Let $P_{k}$ be the vector of all inputs into the system at time $k$, where for $P_{\mathcal{G}_{k}}$  the column vector of generator inputs and $P_{\mathcal{L}_{k} }$ the column vector of load inputs at a time $k$,
 \begin{equation}
P_{k} = \begin{pmatrix} P_{ \mathcal{G}_{k}} \\ P_{ \mathcal{L}_{k} } \end{pmatrix} \, .
\end{equation} 
For matrices $K$ and $B$, the swing dynamics are captured by a discrete linear model of the form
\begin{equation}
\psi(x_{k+1})  = K( ~ \psi(x_{k}) - BP_{k} ~ ) + BP_{k} \, . \label{KO_og_dyn}
\end{equation}
$K$, $B$, and the weights $w$ of $N(x_{k};w)$ are trained with stochastic gradient descent to minimize the model error
\begin{equation}
\dfrac{ || K(  ~ \psi(x_{k}) - BP_{k}~) + BP_{k} -  \psi(x_{k+1}) ||^{2} } { || x_{k} - x_{k+1} ||^{2}   + \epsilon } \, , 
\label{trn_cst}
\end{equation}
where $\epsilon>0$ is taken to be a small value.
Additional constraints are also imposed on $K$ and $B$ such they capture the equilibria and asymptotic behavior of the swing dynamics, full details are in~\cite{king2021}.

\subsection{Dynamics-Aware Economic Dispatch with the Koopman Operator}
Using the KO dynamics \eqref{KO_dyn} in place of the constraints \eqref{swng1} - \eqref{swng4} allows for a linear formulation of the DED problem with the KO dynamics (DED-KO) as follows

\begin{subequations}
\begin{align}
&\min_{P_{i}(t_{k}) \,  i\in \mathcal{G}, k\in \mathcal{T} }  \sum_{k \in \mathcal{T} } \left( \sum_{i \in \mathcal{G} } c_{i} P_{i} \left(t_{k}\right) + c_s |P_{s}(t_{k}) | \right ) \, , \label{KO_obj} \\
&\mbox{subject to:} \nonumber  \\
&\psi_{k+1} = K \left( \psi_{k} - B P \left(t_{k}\right) \right) + B P \left(t_{k}\right) \, , ~ \forall ~ k \in \mathcal{T} \, , \label{KO_dyn} \\
&\psi_{0} = \psi\left(x_0\right) \, , \\
&P_{s}(t_{k})  =  \mathbf{1}_{s}^{T} \psi_{k} \, , ~ \forall ~ k \in \mathcal{T} \, , \\
&\omega_{min} \leq A \psi_{k} \leq \omega_{max} \, , ~ \forall ~ k \in \mathcal{T} \, , \label{KO_ombnd} \\
&P_{i}^{\min} \leq P_{i}(t_{k})  \leq P_{i}^{\max}\, , ~ \forall ~ k \in \mathcal{T} \,  , \, i\in \mathcal{G} \, , \label{KO_gbnd}\\
&  | P_{i}(t_{k+1}) - P_{i}(t_{k}) | \leq \epsilon_{rmp}\, , ~  \forall ~ k \in \mathcal{T} \,  , \, i \in \mathcal{G} \, . \label{KOrmp}.  \\
& P_{i}(t_{0} ) = P_{0_{i}} ~  \forall ~ \, i \in \mathcal{G} \label{KO_ginit}
\end{align}
\label{DED_KO}
\end{subequations}
The matrix $A$ recovers the generator frequencies and has block form $A=[I,{\bf 0}]$ with $I$ the identity matrix corresponding to the position of the generator frequencies in the observable vector $\psi$. Similarly, the vector $\mathbf{1}_{s}$ recovers the value of $P_{s}$ with zeros in all positions except a one at the position corresponding its value in $\psi$.

\section{Dynamics-Aware Economic Dispatch as a Differentiable Predictive Control Problem}

\begin{figure*}[ht!]
  \centering
  \includegraphics[scale=.5]{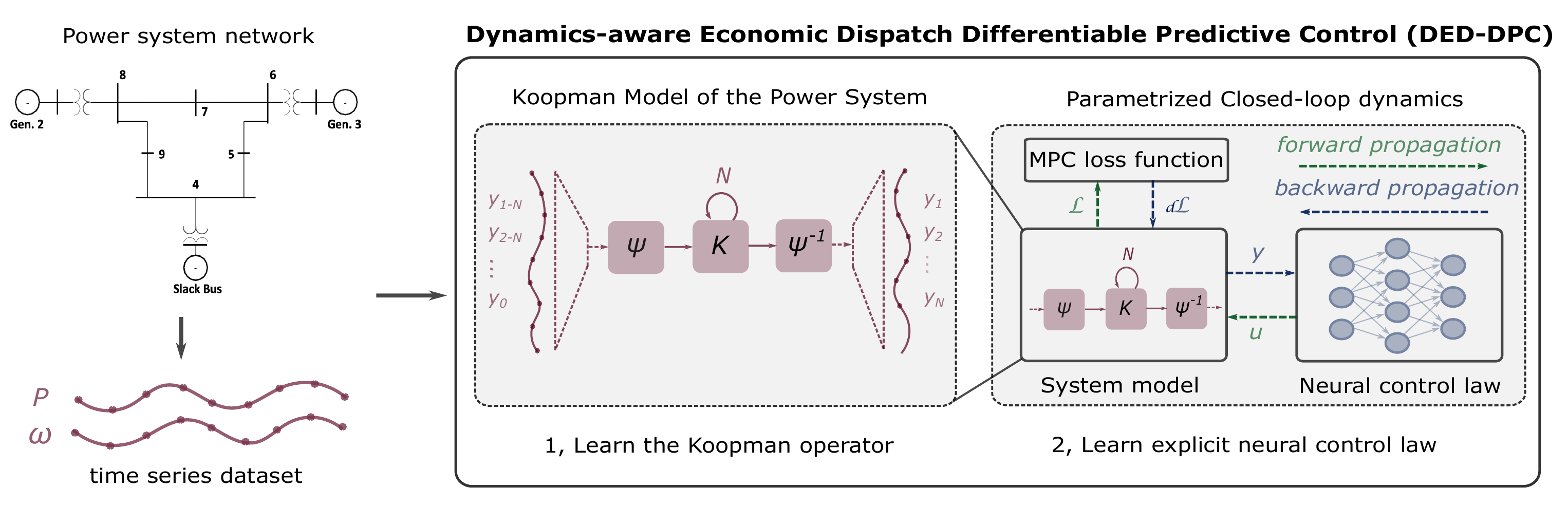}
 \caption{Koopman-based differentiable predictive control (DPC)  approach to dynamics-aware economic dispatch (DED) problem. }
 \label{fig:DED-DPC}
\end{figure*}

 Here we use the KO dynamics as described in~\cite{king2021} as a surrogate model of the swing dynamics in the DED problem formulated and solved via the DPC framework~\cite{drgona2022Gurantees}. The KO  provides a computationally simpler model than the original DAE system that we use  to propagate the affect of control inputs on the state and evaluate them against the MPC objective. Using the KO model we develop a DED differentiable predictive control (DED-DPC) architecture that integrates a predictive neural control policy with the objective and constraints of the DED problem.
The overall methodology is conceptually illustrated in Fig.~\ref{fig:DED-DPC}.

\subsection{Differentiable Predictive Control Problem}

We consider the following general parametric differentiable predictive control (DPC) problem:
\begin{subequations}
\label{eq:DPC}
    \begin{align}
 \min_{{\Theta}} & \ \mathcal{L}_{\text{DPC}} =  \sum_{k=0}^{N-1}  \big( \ell( { x}_k, { u}_k, \zeta_k  )  +  
 \\ &  Q_x ||\sigma(g({ x}_k, \zeta_k)) ||^2_2 +  Q_u ||\sigma(h({ u}_k,  \zeta_k)) ||^2_2 \big) & 
 \label{eq:DPC:objective} \\ 
  \text{s.t.} \ &   { x}_{k+1} = f({ x}_k, { u}_k), \  k \in \mathbb{N}_{0}^{N-1}   \label{eq:dpc:x}  & \\
 \ & { u}_k = \pi(\zeta_k; \Theta)  \label{eq:dpc:pi} \\
  \ & \zeta_k \in \Xi \subset \mathbb{R}^n \label{eq:dpc:xi}
\end{align}
\end{subequations}
Where ${ x}_k$, and ${u}_k$, represent states and control inputs at time $k$, respectively.
The closed loop system dynamics constraints are given explicitly via system model~\eqref{eq:dpc:x} and 
control policy~\eqref{eq:dpc:pi}. While the parametric state  $g({ x}_k, \zeta_k) \le 0$ and control input constraints $h({u}_k, \zeta_k) \le 0$
are penalized in the loss function~\eqref{eq:DPC:objective} via $2$-norm penalty functions, where $\sigma$ represents the rectified linear unit (ReLU) function. The first term of the DPC loss $(\ell { x}_k, { u}_k, \zeta_k)$ represents a parametric objective function, e.g., reference tracking.
The control policy~\eqref{eq:dpc:pi}  maps 
the control parameters $\zeta_k$ to control actions ${ u}_k$, and is parameterized by weights and biases $\Theta$ that need to be optimized. To solve the problem~\eqref{eq:DPC}, we compute the loss function values~\eqref{eq:DPC:objective} over the sampled control parameters $\zeta_k$ from a given set $ \Xi$~\eqref{eq:dpc:xi}. Then we compute the gradient of the loss $\nabla_{\Theta}\mathcal{L}_{\text{DPC}}$ with respect to the policy parameters $\Theta$ and update their values via stochastic gradient descent.
We formalize this procedure in Algorithm~\ref{algo:DPC_optim}.
\begin{algorithm}
  \caption{DPC policy optimization.}\label{algo:DPC_optim}
  \begin{algorithmic}[1]
  \State \textbf{input} training dataset  $\Xi = \{ \zeta_1, \ldots,  \zeta_m\}$
  \State \textbf{input} learned or known system dynamics model ${ x}_{k+1} = f({ x}_k, { u}_k)$
 \State \textbf{input}  policy architecture  $ \pi(\zeta_k; \Theta)$
  \State \textbf{input} DPC loss $\mathcal{L}_{\text{DPC}}$ penalizing performance objective function $\ell({ x}_k, { u}_k, \zeta_k)$, and state and input constraints
 $g({ x}_k, \zeta_k) \le 0$, and $h({u}_k, \zeta_k) \le 0$
 \State \textbf{input} optimizer $\mathbb{O}$
 \State \textbf{train} $\pi(\zeta_k; \Theta)$ to minimize DPC loss $\mathcal{L}_{\text{DPC}}$ using optimizer  $\mathbb{O}$
 \Indent
 \For{$k \gets 1$ to $n_{\text{epochs}}$}
  \State  forward pass the dataset  $\Xi$ through policy  $ \pi(\zeta_k; \Theta)$ and system dynamics ${ x}_{k+1} = f({ x}_k, { u}_k)$
  \State  calculate loss $\mathcal{L}_{\text{DPC}}$ and its gradient  $\nabla_{\Theta}\mathcal{L}_{\text{DPC}}$
  \State update policy parameters $\Theta$ via optimizer $\mathbb{O}$
  \EndFor
 \EndIndent
\State \textbf{return} optimized policy  $ \pi(\zeta_k; \Theta)$ 
  \end{algorithmic}
\end{algorithm}

\subsection{DED-DPC Problem Formulation}

In this section, we show how to convert the original DED problem into the DPC form~\eqref{eq:DPC} to be solved 
via Algorithm~\ref{algo:DPC_optim}.
We start by converting the DED objective \eqref{KO_obj} and constraints \eqref{KO_dyn}-\eqref{KO_ginit}
into the  DPC form~\eqref{eq:DPC}.
The problem input parameters $\zeta$ here represent the loads forecast 
\begin{equation}
    \zeta = ( \{P_{\mathcal{L}} \}_{k \in \mathcal{T}}, c, x_{0}, P_{\mathcal{G}_{0}} )
\end{equation}
where $\{ P_{\mathcal{L}} \}_{k \in \mathcal{T} }$ is the sequence of forecast-loads over the time horizon, $c$ is the vector of the generator cost coefficients, $x_{0}$ is the initial state of the system and $P_{\mathcal{G}_{0}}$ is the vector of initial generator inputs. 
The outputs of the control policy is a sequence of generator inputs, where we denote the input to the $i$-th generator at time $k$ as follows
\begin{equation}
P_{i_{k}} = \pi_{i_{k}}(\zeta;\Theta)  \,.
\end{equation}
To evaluate the control policy we construct the model sequence
$\{\psi_{\pi_{k}}(\zeta)\}_{k \in \mathcal{T}}$ subject to $~\forall~ k \in \mathcal{T}$\,
{\small
\begin{subequations}
\begin{align}
    & \psi_{\pi_{0}}(\zeta) = \psi(x_{0}) \\
    &\psi_{\pi_{k+1}}(\zeta) = K( ~ \psi_{\pi_{k}}(\zeta) - BP_{k} ~ ) + BP_{k} \\
    &P_{k} = \begin{pmatrix} \pi_{k}(\zeta;\Theta) \\ P_{ \mathcal{L}_{k} } \end{pmatrix} \, .
\end{align}
\end{subequations}
}

Now given a set of paramters $\mathcal{J}$ and by leveraging the penalty constraints we obtain the following differentiable DED-DPC loss $\mathcal{L}_{\text{DED-DPC}}$:
{\small
\begin{subequations}
\begin{align}
    \frac{1}{|\mathcal{J}|} \sum_{j\in \mathcal{J} }  &\sum_{k \in \mathcal{T} } \left( \sum_{i \in \mathcal{G} } c_{i} \pi_{i,k}(\zeta_{j}) + c_s \left|\mathbf{1}_{s}^{T} \psi_{\pi_{k}}(\zeta_{j}) \right| \right ) \label{loss_obj} \\
     +& \sum_{k \in \mathcal{T} } Q_{\omega}\left| \left| \sigma( A \psi_{\pi_{k}}(\zeta_{j}) - \boldsymbol \omega_{max} ) \right| \right|^{2} \label{loss_omu} \\
     +& \sum_{k \in \mathcal{T} } Q_{\omega}\left| \left| \sigma( - A \psi_{\pi_{k}}(\zeta_{j}) + \boldsymbol \omega_{min} ) \right| \right|^{2}\label{loss_oml}  \\
     +& \sum_{k \in \mathcal{T} }Q_{r}\left| \left| \sigma (\pi_{k}(\zeta_{j}; \Theta) -\pi_{k+1}(\zeta_{j};\Theta) - \boldsymbol \epsilon_{rmp} ) \right| \right|^{2} \label{loss_rd} \\
     +& \sum_{k \in \mathcal{T} }Q_{r}\left| \left| \sigma (\pi_{k+1}(\zeta_{j};\Theta) -\pi_{k}(\zeta_{j};\Theta) - \boldsymbol \epsilon_{rmp} ) \right| \right|^{2} \label{loss_ru} \\
     +&  ~Q_{0}\left| \left| \pi_{0} (\zeta_{j}) -P_{\mathcal{G}_{0}} \right| \right|^{2} \, , \label{loss_ginit} 
\end{align}
\label{DPC_loss}
\end{subequations}
}
To train a control policy, for a set of parameters $\{\zeta_{j}\}_{j \in \mathcal{J}}$ sampled out of the expected operating conditions, we use stochastic gradient descent to minimize the differentiable DED-DPC loss~\eqref{DPC_loss}.
The formulation of the DPC loss mirrors the DED-KO formulation \eqref{DED_KO}. The first term in the loss \eqref{loss_obj} is the DED-KO objective \eqref{KO_obj} while the remaining terms enforce the state constraints \eqref{KO_ombnd}, \eqref{KOrmp}, and \eqref{KO_ginit}. In particular, the loss terms \eqref{loss_oml} and \eqref{loss_omu} penalize control policies which result in violations of the generator frequency bounds \eqref{KO_ombnd}. If the frequencies remain within the bounds the terms will evaluate to zero, however for time-steps in which the bounds are crossed they add a penalty proportional to the size of the violation and weighted by the parameter $Q_{\omega}$. Therefore by minimizing the loss, control policies will be adjusted to maintain the state within the bounds. Similarly, the loss terms \eqref{loss_rd}-\eqref{loss_ru} capture the generator ramping constraints \eqref{KOrmp} and the initial generator input constraint \eqref{KO_ginit} is captured by the term \eqref{loss_ginit}.

\subsection{Control Policy Architecture with Hard Input Constraints}
\label{cntrl_policy}

Note that the control upper and lower bound constraints \eqref{KO_gbnd} are not represented in the loss, instead they are enforced as hard constraints by constructing the control policy output such that they are always satisfied. In general within DPC approaches, hard constraints may be imposed on free variables but satisfying constraints on dependent variables must be learned through the selection of appropriate free variables by imposing penalties in the loss function.    

In this paper, we learn a control policy mapping the problem parameters  $\zeta$ to a schedule of power generation with a three layer convolutional neural network (CNN) architecture. For a given number of input channels $n_{in}$, output channels $n_{out}$ and window size $w_{d}$, we use $n_{out}$ one-dimensional convolutional layers with kernels of dimension $(n_{in},w)$ convolved along the time dimension with stride one and padded with zeros such that the dimension is not reduced.

To learn the policy, we sample the DED problem on a minute time interval, while the dynamics of the system are modeled on a $0.01$ second time scale. Therefore both the load trajectory inputs and generator control policies require $6000$ time steps. However, the loads and corresponding controls change relatively slowly over the time horizon. To reduce the complexity of the problem we reduce the dimension using a down-sampled timescale denoted $\mathcal{T}_{d}$ that selects $N_{d}$ evenly spaced time-points.
The optimal control for DED depends on the forecasted loads as well as the initial state of the system and the cost coefficients for each generator. Therefore as input to the first convolutional layer we take for problem parameters $\zeta$ the sequence on the down-sampled timescale
\begin{equation}
\zeta = \left\{  P_{\mathcal{L}_{k}}, c,  x_{0}, P_{\mathcal{G}_{0}} \right\}_{k \in \mathcal{T}_{d}}
\end{equation}
such that for a number of load nodes $n_{L}$ and additional parameters $n_{p}$, the dimension of the input to the first CNN layer is a sequence with $n_{L}+n_{p}$ channels of length $N_{d}$. 
We use \texttt{ReLU} activation functions for the first two CNN layers, while the output of the last layer is scaled using a hyperbolic tangent activation function to enforce the constraints \eqref{KO_gbnd} on the generator inputs. Specifically, for $C_{3}$ the output of the last CNN, the generator inputs are computed on the down-sampled time-scale $\forall~ k\in \mathcal{T}_{d}$ and $ i \in \mathcal{G}$ according to
\begin{equation}
    \pi_{k_{i}}(\zeta; \Theta ) = P_{i}^{\min} + \frac{1}{2}( 1 + \tanh (C_{3_{i}}))(P_{i}^{\max} - P_{i}^{\min} )
\end{equation}
Generator inputs on the full timescale $\mathcal{T}$ are then computed by linear interpolation.

\section{Case Study on 9 Bus Test System}

We evaluate DED-DPC on a 9-bus system. Parameters for the system were taken from~\cite{king2021}. The transmission line parameters are in Table  \ref{tparm} and the bus parameters are in Table \ref{busparm}, where $P_{n_{i}}$ and $L_{n_{i}}$ are the nominal value for bus $i$. All parameter values are in per unit (pu) unless otherwise indicated. Computations are done with the generator frequencies in radians per second, but results are reported in Hertz. 

\begin{table}[]
 \centering
  \caption{Bus Parameters}
  \label{busparm}
  \begin{tabular}{ccccc}
    \toprule
   Bus & $P_{n_{i}}/L_{n_{i}}$ &  $|V_{i}|$ &  $M_{i}$ & $D_{i}$  \\
    \midrule
    s &  NA  &  1.04 & 13.64 & 9.6 \\
    2 &  1.63  & 1.02533 & 6.4  & 2.5 \\
    3 &  0.85 & 1.02536 & 3.01 & 1.0  \\
    5 &  -0.9 & 1  & NA  & NA \\
    7 &  -1  &  1 & NA & NA  \\
    9 &  -1.25 & 1 & NA  & NA \\
     \bottomrule
\end{tabular}
\end{table}

\begin{table}[]
 \centering
  \caption{Transmission line parameters}
  \label{tparm}
  \begin{tabular}{cccc}
    \toprule
    From bus&To bus&Resistance  &Reactance \\
    \midrule
    s&4& 0 & 0.0576\\
    4&5& 1.7e-5 & 0.092 \\
    5&6& 3.9e-5 & 0.17 \\
    3&6& 0 & 0.0586 \\
    6&7& 1.2e-5 & 0.1008 \\
    7&8& 8.5e-6 & 0.072 \\
    8&2& 0 & 0.0625 \\
    8&9& 3.2e-5 & 0.161 \\
    9&4&1e-5 & 0.085\\
  \bottomrule
\end{tabular}
\end{table}

\subsection{Economic Dispatch Problem Parameters}

Following the work in~\cite{king2021} we test DED-DPC in two operational regimes with different bounds on the generator frequencies. The bounds on the deviation from the reference frequency are taken to be symmetric with $\omega_{bnd} = \omega_{\max} = - \omega_{\min}$. The first regime uses the nominal operating bounds $\omega_{bnd} = 0.05 Hz$ taken from~\cite{Kirby2002}  with the reference frequency assumed to be at $60 Hz$. As was found in~\cite{king2021} the DED solutions easily stay within these frequency bounds and they are never active.
To asses the limits of the DED-DPC under more extreme conditions and compare it to the previous work using DED-KO, we test it also in a tight operating (TO) regime. In this regime the frequency bounds are reduced over an order of magnitude, with $\omega_{bnd} = 0.0016 Hz$.
The control is also required to correct an initial perturbation given by
\begin{equation}
P_{i}(0) = P_{opt_{i}} - 0.5(P_{opt_{i}} - P_{n_{i}})s_{i} \, , ~  \forall ~ i\in \mathcal{G} \, ,
\end{equation}
with $P_{opt_{i}}$ the optimal generator set-point for the initial system loads at generator $i$ and $s_{i}$ a random variable.

The load forecasts the control must match are randomly generated for the nodes $(5,7,9)$ with
{\small
\begin{equation}
P_{i}(t_{k}) = \begin{cases} P_{0_{i}} & t_{k} \leq t_{0_{i}} \\  P_{0_{i}} + r_{i}(t_{k} - t_{0_{i}})   &  t_{0_{i}} \leq t_{k} \leq t_{0_{i}} + d_{i} \\ P_{0_{i}} + r_{i}d_{i} & t_{k} \geq t_{0_{i}} + d_{i} \end{cases}  \end{equation}
}
where $P_{0_{i}}$ is the random initial load and $t_{0_{i}}$, $d_{i}$, and $r_{i}$ are random variables that parameterize the trajectory. 
The random variables associated with initial generator inputs and the load trajectories are all uniformly sampled, along with the initial generator frequencies and the generator cost coefficients, with sampling ranges given in Table \ref{rnd_parms}. The load ramping parameter $r_{i}$ can also be positive or negative with probability $.5$ and can be zero with probability $.15$.

The initial voltage angles are set such that the slack bus and generator inputs exactly satisfy the initial loads. The value of the remaining parameters are given in Table \ref{fxd_parms}.
Over the space of input parameters we generate 500 DED problem instances including initial conditions and load trajectories in both of the NO and TO operating regimes, with 400 problems randomly selected for training the DED-DPC architecture and the other 100 problems for testing.
\begin{table}[]
\renewcommand{\arraystretch}{1.5} 
\centering
  \caption{Dynamic Economic Dispatch Random Problem Parameter Sampling Ranges}
  \begin{tabular}{ll}
    \toprule
    Parameter & Range \\
    \midrule
    $P_{0_{i}}\, , ~  \forall i\in \mathcal{L}$ &  $L_{n_{i}} \pm 0.25 L_{n_{i}} $ \\
    
    $t_{0_{i}}$ & $[0,60]$ \\
     
    $d_{i}$ & $[5,20]$ \\
     
    $r_{i}$ & $\pm[0.01,0.05]$ \\
     
    $s_{i} \, , ~  \forall i \in \mathcal{G} $ & $[0,1]$ \\
     
    $\omega_{0_{i}} \, , ~  \forall i \in \mathcal{G} $ & $[-0.01,0.01] $ \\
     
    $c_{i} \, , ~ \forall i \in \mathcal{G} \cup \{s\} $ & $[0,1]$\\
  \bottomrule
\end{tabular}
\label{rnd_parms}
\end{table}

\begin{table}[]
\renewcommand{\arraystretch}{1.5} 
\centering
  \caption{Dynamic Economic Dispatch Fixed Problem Parameters}
  \begin{tabular}{ll}
    \toprule
    Parameter & Value \\
    \midrule
    $P_{i}^{\min} \, \forall i \in \mathcal{G} $ & $P_{n_{i}} - 0.5P_{n_{i}}$\\
   
    $P_{i}^{\max} \, \forall i \in \mathcal{G} $ & $P_{n_{i}} + 0.5P_{n_{i}}$\\
    
    $\epsilon_{rmp}$ & 5e-4 \\
  \bottomrule
\end{tabular}
\label{fxd_parms}
\end{table}

\subsection{Differentiable Predictive Control Implementation}

We implemented the DED-DPC problem in NeuroMANCER~\cite{Neuromancer2021} which is a differentiable programming library for solving constrained optimization problems developed on top of the Pytorch~\cite{paszke2019pytorch} deep learning library.
For training the DED-DPC control policy in both of the TO and NO operating regimes we use the CNN policy as outlined in Section \ref{cntrl_policy}. We set the down-sampled timescale to have dimension $N_{d} = 50$. For all CNN layers $n_{out}=30$, except for the final layer for which $n_{out}=|\mathcal{G}|$. In the first and last CNN layers $w_{d} = 5$, while for the middle layer $w_{d} = 10$.
The weights for the KO observable map given by neural network $N(,:w)$ as well as the matrices $K$ and $B$ for the KO dynamics \eqref{KO_og_dyn} were taken from~\cite{king2021}. 
We train the control to minimize the loss \eqref{DPC_loss} using stochastic gradient descent with the AdamW optimizer~\cite{loshchilov2017decoupled} and learning rate 5e-4. The weights in the loss for the constraint penalty terms are set as follows: $Q_{\omega}$=1e3, $Q_{r}$=1e2, $Q_{0}$=1e4. Computation was done on an RTX 2080 Ti GPU with 384GB memory and took approximately 3 hours in each regime. 

\subsection{Benchmark Online Optimization Approaches}
For comparison we also solve the full DED \eqref{DED_OG} and DED-KO \eqref{DED_KO} optimization problems in python using pyomo~\cite{hart2011pyomo}. For DED the swing dynamic constraints are enforced with pyomo's DAE solver~\cite{pyomoDAE}, while both DED constrained optimization problem formulations are solved using IPOPT~\cite{WachterIpopt} with the default tolerances. Computations was done on a Macbook Pro with a 2.3 GHz 8-Core Intel Core i9 processor and 16 GB 2667 MHz DDR4 memory.

\section{Results}

To evaluate and compare the performance of the controls we simulate the swing dynamics \eqref{swng_gen}-\eqref{swng_f} under the control actions from each method using MATLAB's algebraic differential equation solver ode15i. All constraint violation and objective evaluations are done with these simulations.

\begin{figure}[h]
  \centering
  \includegraphics[scale=.1]{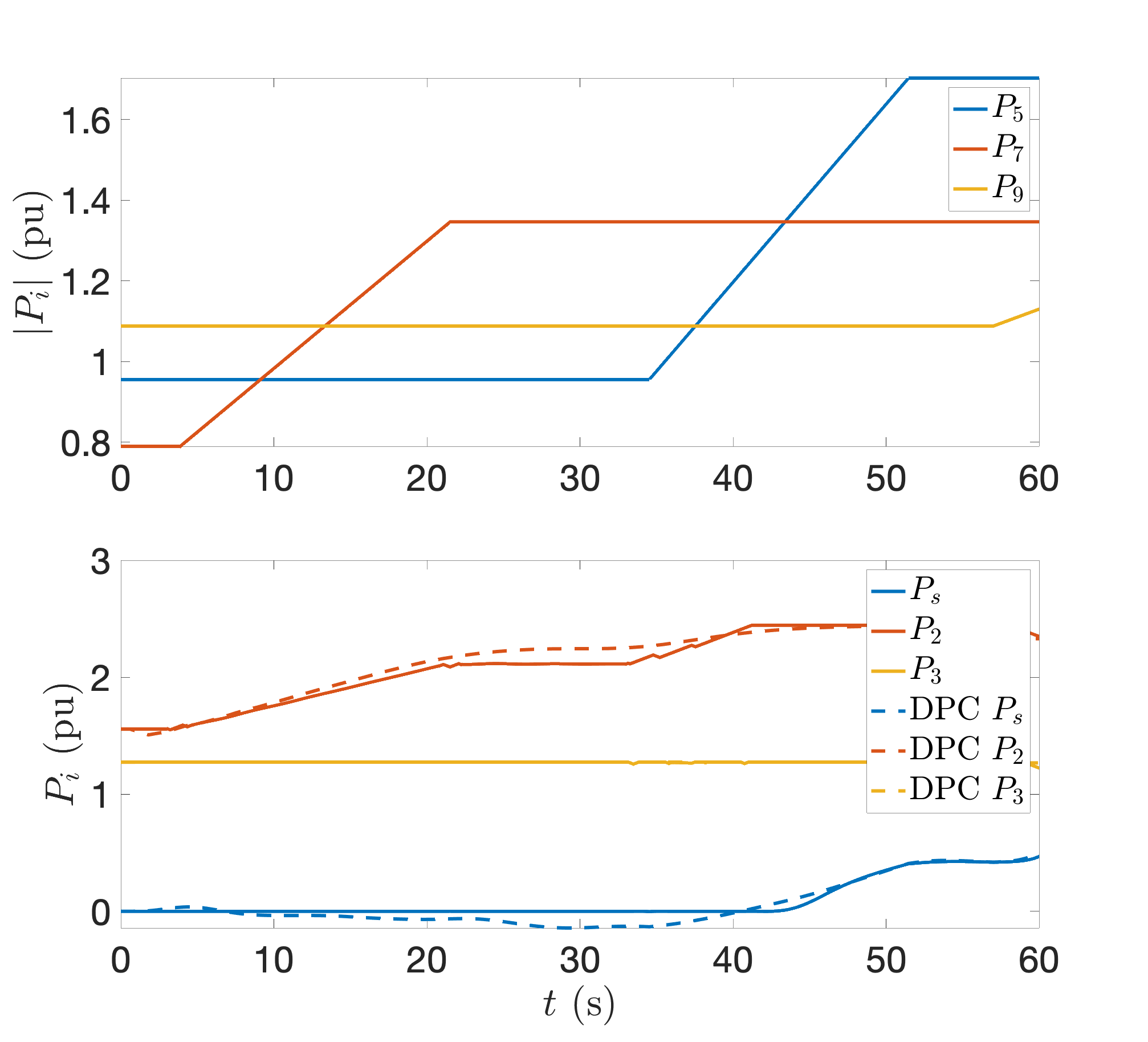}
 \caption{Comparison of a differential parametric control (DPC) solution and IPOPT solution for a  nominal operating (NO) regime test problem.}
 \label{NO_cmp}
\end{figure}

\begin{figure}[]
  \centering
  \includegraphics[scale=.22]{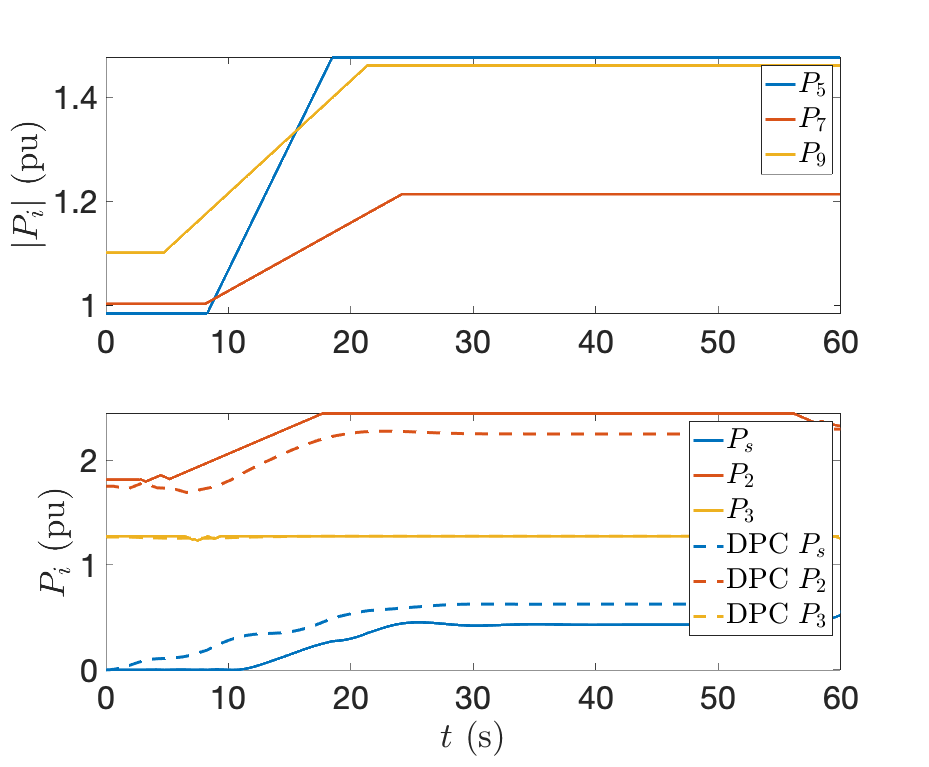}
 \caption{Comparison of a differential parametric control (DPC) solution and IPOPT solution for a tight operating (TO) regime test problem.}
 \label{TO_cmp}
\end{figure}

As shown in Table \ref{oandt}, at implementation the DED-DPC method generates solutions $5$ orders of magnitude faster than either the DED or DED-KO approach. In the nominal operating regime the DED-DPC method closely recovers the DED solutions with only a $1$\% average increase in the generation cost over one hundred test problems. Figure \ref{NO_cmp} shows the close match between a DED-DPC and DED solution on a representative NO test problem. Some small ramping constraint violations are also introduced in the NO regime when using the DED-DPC solutions as given Table \ref{cns_viol} but are negligible on average.
Even in the tight operating regime the DED-DPC solutions do well to match the loads and stay within state constraints. As shown in Figure \ref{TO_cmp} the DED-DPC solutions respond more conservatively to changes in the load in the TO regime, likely the control policy is relying more on the slack bus to match variation in load as a means to better regulate the generator frequencies, which results in higher average generation costs.

\begin{figure}[]
  \centering
  \includegraphics[scale=.22]{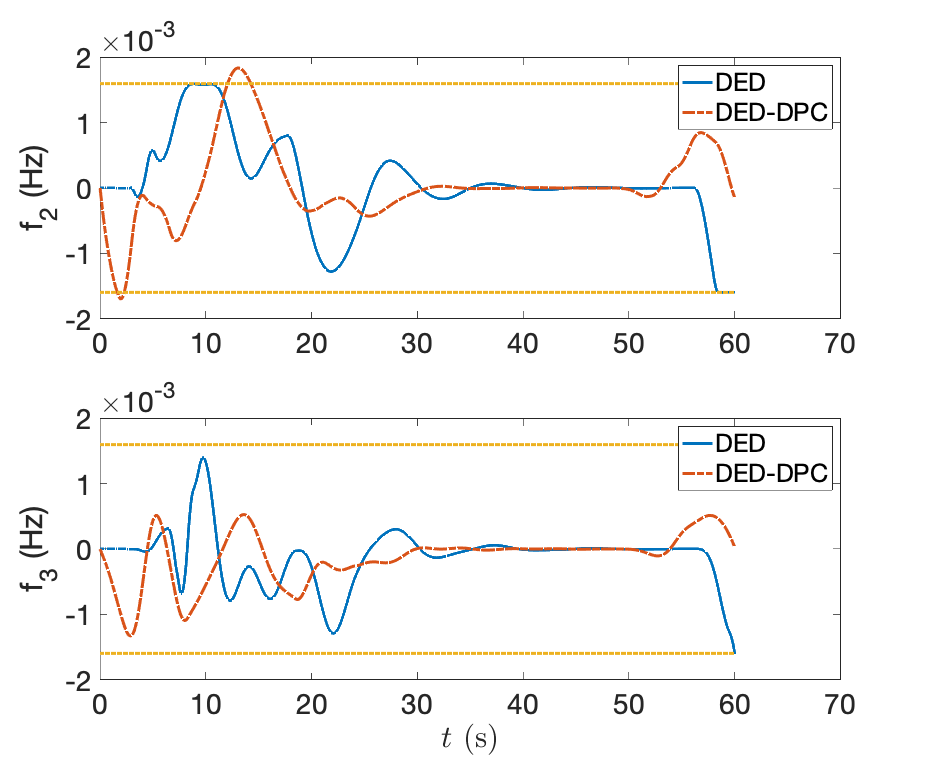}
 \caption{Comparison between the generator frequencies under control from the differential parametric solution (DED-DPC) and control from the IPOPT interior point solution (DED) in the tight operating (TO) regime, the dotted lines are the the frequency bounds}
 \label{freq_cmp}
\end{figure}

\begin{figure}[htb]
  \centering
  \includegraphics[scale=.22]{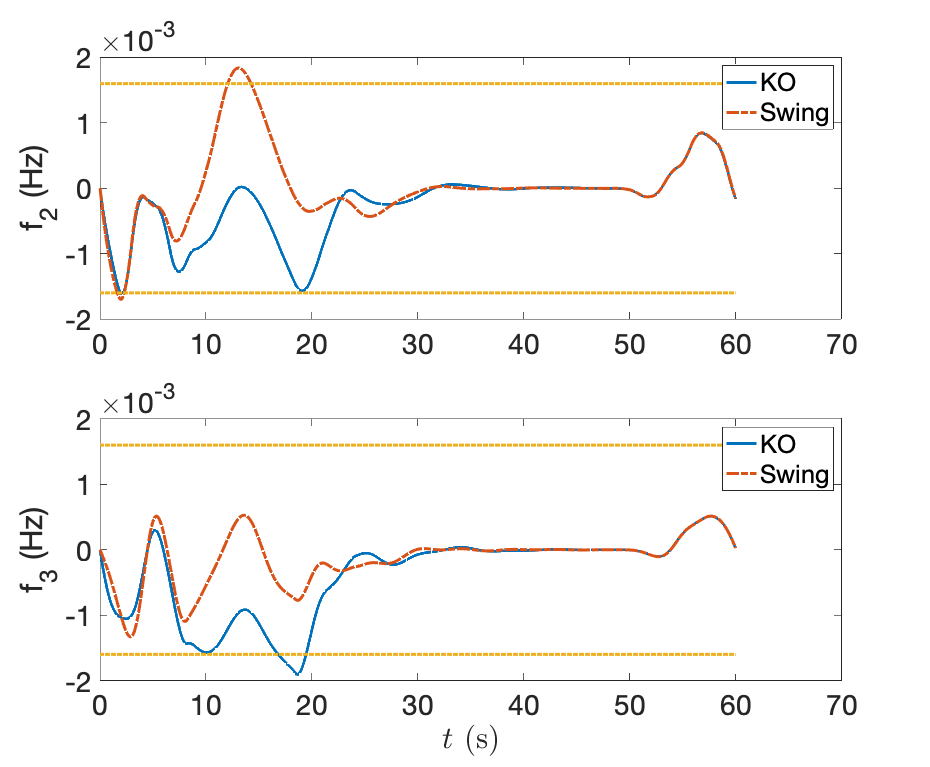}
 \caption{Comparison between the generator frequency trajectories when simulating the system under the differentiable parametric control solution with either the Koopman model or with the swing equations.}
 \label{KOswng_cmp}
\end{figure}

Generator frequency violations are observed for the DED-DPC solutions in the TO regime, though as shown Figure \ref{freq_cmp} they are generally small and of short duration. Interestingly, as given in Table \ref{cns_viol}, the average maximum value of these violations is roughly the same as what occur in the DED-KO solutions. A potential large contributor to these violations is mismatch between the KO model and swing dynamics. Figure \ref{KOswng_cmp} shows a comparison between the generator frequency trajectories when the DED-DPC control inputs are simulated in the KO model versus the swing dynamic model. The observed frequency violation coincides with a deviation in the trajectories where there is a sharp increase in the loads on the system. Note that in the KO model simulation there is also a constraint violation but in the generator 3 frequency. Changing the weight $Q_{\omega}$ on the penalty term in the DED-DPC loss function \eqref{DPC_loss} could be used to try to correct these KO model violations but the model mismatch would likely still result in violations for the swing dynamics. This suggests further improvement in performance in the TO regime would require improving the KO model approximation.

\begin{table}[]
\centering
  \caption{Average Solution Time and Change in Objective}
 \begin{tabular}{clcc}
    \toprule
    Regime & Method & \% increase in obj. over DED & Sol. time (s) \\
    \midrule
     NO & DED & NA &  32.34\\
        & DED-KO & 0.094 &  17.34     \\
       & DED-DPC & 1.05 & 6.8e-4 \\
    TO & DED  &  NA & 32.38\\
       & DED-KO &  0.3   & 18.8          \\
       & DED-DPC & 5.2 & 6.8e-4 \\
   
  \bottomrule
\end{tabular}
\label{oandt}
\end{table}

\begin{table}[]
\centering
  \caption{Average maximum constraint violation over Test Problems}
  \begin{tabular}{clcc}
    \toprule
    Regime & Method & freq. const. viol. & ramp cont. viol. \\
    \midrule
     NO & DED & 0 &  1.84e-9\\
        & DED-KO & 0 & 2.0e-9    \\
       & DED-DPC & 0 & 1.11e-4 \\
    TO & DED  & 7.5e-5 & 1.9e-9\\
       & DED-KO & 0.0013  & 1.9e-9    \\
       & DED-DPC & 0.0011 & 4.2e-5 \\
   
  \bottomrule
\end{tabular}
\label{cns_viol}
\end{table}

\subsection{Scalability of the DPC Method}
The DED-DPC approach performs well on the 9-bus system in this study but must be able to scale in order to be feasible for larger systems of more practical relevance. The largest computational bottleneck for DED-DPC is in training the parametric control policy offline. As the problem size and size of the parameter space increases the training cost may increase significantly. However, once trained, online computational time is dictated only by the neural control policy architecture and in general will remain relatively fast. On the 9-bus system training took about three hours. One of the biggest contributors to the computational time is the need to simulate out the full system dynamics (6000 time-steps for the problem here) for every parameter input evaluated in a training step. To lessen this burden, as suggested in~\cite{king2021}, the KO could potentially be used to reduce the model dimension for larger grids. The KO could also be leveraged to learn a dynamic model at a coarser time-scale to further reduce the computational requirements for simulation during training of larger systems, though at the potential cost of reduced accuracy.

\section{Conclusions and Future Work}

In this paper, we have presented a new solution approach to the dynamics-aware economic dispatch problem based on differentiable predictive control (DED-DPC).
Compared to traditional online constrained optimization solvers, the presented DED-DPC method obtains the parametric solution offline. In the simulation case study using a 9-bus system, we demonstrate that it can provide up to five orders of magnitude speed-ups in online computation speed.
The proposed approach maintains high solution quality with only marginal increases in generation cost while closely satisfying generator frequency constraints. Recently, a method for probabilistic stability and constraint satisfaction guarantees was presented in~\cite{drgona2022Gurantees} for DPC approaches that we will explore for our DED implementation in future work. We will also explore the scalability of the proposed DED-DPC approach through data and model parallelism and increase in GPU computational resources to deal with practical problems in large-scale power system networks.










\bibliographystyle{plain}
\bibliography{bib}

\begin{thebibliography}{10}

\bibitem{Shri2017}
Shrirang Abhyankar, Guangchao Geng, Mihai Anitescu, Xiaoyu Wang, and Venkata
  Dinavahi.
\newblock Solution techniques for transient stability-constrained optimal power
  flow – part i.
\newblock {\em IET Generation, Transmission \& Distribution}, 11:3177--3185(8),
  August 2017.

\bibitem{diffMPC2018}
Brandon Amos, Ivan Dario~Jimenez Rodriguez, Jacob Sacks, Byron Boots, and
  J.~Zico Kolter.
\newblock Differentiable {MPC} for end-to-end planning and control.
\newblock {\em CoRR}, abs/1810.13400, 2018.

\bibitem{Arapostathis82}
Aristotle Arapostathis, S.~Shank Sastry, and Varaiya Pravin.
\newblock Global analysis of swing dynamics.
\newblock {\em IEEE Transaction on Circuits and Systems}, 29(10):673--678,
  1982.

\bibitem{bakker20cp}
Craig Bakker, Arnab Bhattacharya, Samrat Chatterjee, Casey~J Perkins, and
  Matthew~R Oster.
\newblock The koopman operator: Capabilities and recent advances.
\newblock In {\em 2020 IEEE Resilience Week (RWS)}, 2020.

\bibitem{budisic12jsr}
Marko Budi{\v{s}}i{\'c}, Ryan Mohr, and Igor Mezi{\'c}.
\newblock Applied koopmanism.
\newblock {\em Chaos: An Interdisciplinary Journal of Nonlinear Science},
  22(4):047510, 2012.

\bibitem{chakraborty2020}
P.~{Chakraborty}, S.~{Dhople}, C.~{Yu Chen}, and M.~{Parvania}.
\newblock Dynamics-aware continuous-time economic dispatch and optimal
  automatic generation control.
\newblock In {\em 2020 American Control Conference (ACC)}, pages 1292--1298,
  2020.

\bibitem{GNURL2019}
Bingqing Chen, Zicheng Cai, and Mario Berg\'{e}s.
\newblock Gnu-rl: A precocial reinforcement learning solution for building hvac
  control using a differentiable mpc policy.
\newblock In {\em Proceedings of the 6th ACM International Conference on
  Systems for Energy-Efficient Buildings, Cities, and Transportation}, BuildSys
  '19, page 316–325, New York, NY, USA, 2019. Association for Computing
  Machinery.

\bibitem{Chen2018}
S.~{Chen}, K.~{Saulnier}, N.~{Atanasov}, D.~D. {Lee}, V.~{Kumar}, G.~J.
  {Pappas}, and M.~{Morari}.
\newblock Approximating explicit model predictive control using constrained
  neural networks.
\newblock In {\em 2018 Annual American Control Conference (ACC)}, pages
  1520--1527, June 2018.

\bibitem{drgona2020differentiable}
Jan Drgona, Karol Kis, Aaron Tuor, Draguna Vrabie, and Martin Klauco.
\newblock {Differentiable Predictive Control: An MPC Alternative for Unknown
  Nonlinear Systems using Constrained Deep Learning}.
\newblock arXiv:2011.03699, 2020.

\bibitem{drgona2022Gurantees}
Jan Drgona, Aaron Tuor, and Draguna Vrabie.
\newblock Learning constrained adaptive differentiable predictive control
  policies with guarantees.
\newblock arXiv:2004.11184, 2022.

\bibitem{DRGONA2018}
J\'an Drgoňa, Damien Picard, Michal Kvasnica, and Lieve Helsen.
\newblock Approximate model predictive building control via machine learning.
\newblock {\em Applied Energy}, 218:199 -- 216, 2018.

\bibitem{DRGONA202114}
Ján Drgoňa, Aaron Tuor, Elliott Skomski, Soumya Vasisht, and Draguna Vrabie.
\newblock Deep learning explicit differentiable predictive control laws for
  buildings.
\newblock {\em IFAC-PapersOnLine}, 54(6):14--19, 2021.
\newblock 7th IFAC Conference on Nonlinear Model Predictive Control NMPC 2021.

\bibitem{east2020infinitehorizon}
Sebastian East, Marco Gallieri, Jonathan Masci, Jan Koutnik, and Mark Cannon.
\newblock Infinite-horizon differentiable model predictive control.
\newblock arXiv:2001.02244, 2020.

\bibitem{hart2011pyomo}
William~E Hart, Jean-Paul Watson, and David~L Woodruff.
\newblock Pyomo: modeling and solving mathematical programs in python.
\newblock {\em Mathematical Programming Computation}, 3(3):219--260, 2011.

\bibitem{Hertneck8371312}
M.~{Hertneck}, J.~{Köhler}, S.~{Trimpe}, and F.~{Allgöwer}.
\newblock Learning an approximate model predictive controller with guarantees.
\newblock {\em IEEE Control Systems Letters}, 2(3):543--548, 2018.

\bibitem{DiffProg2019}
Mike Innes, Alan Edelman, Keno Fischer, Christopher Rackauckas, Elliot Saba,
  Viral~B. Shah, and Will Tebbutt.
\newblock A differentiable programming system to bridge machine learning and
  scientific computing.
\newblock {\em CoRR}, abs/1907.07587, 2019.

\bibitem{KARG2021107266}
Benjamin Karg and Sergio Lucia.
\newblock Approximate moving horizon estimation and robust nonlinear model
  predictive control via deep learning.
\newblock {\em Computers \& Chemical Engineering}, 148:107266, 2021.

\bibitem{khatami20}
Roohallah Khatami, Masood Parvania, Swaroop Guggilam, Christine Chen, and
  Sairaj Dhople.
\newblock Dynamics-aware continuous-time economic dispatch: A solution for
  optimal frequency regulation.
\newblock In {\em Proceedings of the 53rd Hawaii International Conference on
  System Sciences}, pages 3186--3195, Honolulu, HI, 01 2020.

\bibitem{king2021}
Ethan King, Craig Bakker, Arnab Bhattacharya, Samrat Chatterjee, Feng Pan,
  Matthew~R Oster, and Casey~J Perkins.
\newblock Solving the dynamics-aware economic dispatch problem with the koopman
  operator.
\newblock In {\em The Twelfth ACM International Conference on Future Energy
  Systems}. ACM, 2021.

\bibitem{Kirby2002}
B.J. Kirby, C.~Martinez, J.~Dyer, A.~Shoureshi, D.~Rahmat, J.~Dagle, and
  R.~Guttromson.
\newblock Frequency control concerns in the north american electric power
  system.
\newblock Technical report, Oak Ridge National Laboratory, 2002.

\bibitem{Khler2017RealTE}
Johannes K{\"o}hler, Matthias~Albrecht M{\"u}ller, Na~Li, and Frank
  Allg{\"o}wer.
\newblock Real time economic dispatch for power networks: A distributed
  economic model predictive control approach.
\newblock {\em 2017 IEEE 56th Annual Conference on Decision and Control (CDC)},
  2017.

\bibitem{lee2013}
Y.~{Lee} and R.~{Baldick}.
\newblock A frequency-constrained stochastic economic dispatch model.
\newblock {\em IEEE Transactions on Power Systems}, 28(3):2301--2312, 2013.

\bibitem{li2016}
N.~{Li}, C.~{Zhao}, and L.~{Chen}.
\newblock Connecting automatic generation control and economic dispatch from an
  optimization view.
\newblock {\em IEEE Transactions on Control of Network Systems}, 3(3):254--264,
  2016.

\bibitem{loshchilov2017decoupled}
Ilya Loshchilov and Frank Hutter.
\newblock Decoupled weight decay regularization.
\newblock {\em arXiv preprint arXiv:1711.05101}, 2017.

\bibitem{maddalena2019neural}
E.~T. Maddalena, C.~G. da~S.~Moraes, G.~Waltrich, and C.~N. Jones.
\newblock A neural network architecture to learn explicit mpc controllers from
  data, 2019.

\bibitem{milano2018}
F.~{Milano}, F.~{Dörfler}, G.~{Hug}, D.~J. {Hill}, and G.~{Verbič}.
\newblock Foundations and challenges of low-inertia systems (invited paper).
\newblock In {\em 2018 Power Systems Computation Conference (PSCC)}, pages
  1--25, 2018.

\bibitem{pyomoDAE}
Bethany Nicholson, John~D. Siirola, Jean-Paul Watson, Victor~M. Zavala, and
  Lorenz~T. Biegler.
\newblock pyomo.dae: a modeling and automatic discretization framework for
  optimization with differential and algebraic equations.
\newblock {\em Mathematical Programming Computation}, 2018.

\bibitem{paszke2019pytorch}
Adam Paszke, Sam Gross, Francisco Massa, Adam Lerer, James Bradbury, Gregory
  Chanan, Trevor Killeen, Zeming Lin, Natalia Gimelshein, Luca Antiga, et~al.
\newblock Pytorch: An imperative style, high-performance deep learning library.
\newblock In {\em Advances in Neural Information Processing Systems}, pages
  8024--8035, 2019.

\bibitem{thatte2011}
A.~A. {Thatte}, {Fan Zhang}, and L.~{Xie}.
\newblock Frequency aware economic dispatch.
\newblock In {\em North American Power Symposium}, pages 1--7, 2011.

\bibitem{trip2016}
Sebastian Trip, Mathias Bürger, and Claudio {De Persis}.
\newblock An internal model approach to (optimal) frequency regulation in power
  grids with time-varying voltages.
\newblock {\em Automatica}, 64:240 -- 253, 2016.

\bibitem{Neuromancer2021}
Aaron Tuor, Jan Drgona, and Elliot Skomski.
\newblock {NeuroMANCER: Neural Modules with Adaptive Nonlinear Constraints and
  Efficient Regularizations}.
\newblock 2021.

\bibitem{WachterIpopt}
A.~W\"{a}chter and L.T. Biegler.
\newblock On the implementation of a primal-dual interior point filter line
  search algorithm for large-scale nonlinear programming.
\newblock {\em Mathematical Programming}, 106(1):25--57, 2006.

\bibitem{wood2013}
A.J. Wood, B.F. Wollenberg, and G.B Shebl'e.
\newblock {\em Power Generation, Operation, and Control (3rd edition)}.
\newblock John Wiley, New York, 2013.

\bibitem{safe_RL_MPC_2019}
Mario Zanon and S{\'{e}}bastien Gros.
\newblock Safe reinforcement learning using robust {MPC}.
\newblock {\em CoRR}, abs/1906.04005, 2019.

\end{thebibliography}

\end{document}